\documentstyle[preprint,amsfonts,amssymb,epsfig,aps]{revtex}
\tightenlines
\newcommand{\rv}{{\bf r}}

\newcommand{\half}{{1\over 2}}

\newcommand\eqref[1]{Eq.~(\ref{#1})}
\newcommand\klref[1]{(\ref{#1})}
\newcommand\cref[1]{Ref.~\cite{#1}}
\newcommand\abbref[1]{Fig.~\ref{#1}}
\newcommand\eqpref[1]{Eq.~(\protect\ref{#1})}
\newcommand\abbpref[1]{Fig.~\protect\ref{#1}}
\newcommand\pref[1]{\protect\ref{#1}}

\newcommand{\nn}{\nonumber}

\newcommand\rhos{\rho^\ast}

\begin{document}
\draft
\title{Vesicles in solutions of hard rods}
\author{B. Groh}
\address{
  FOM Institute for Atomic and Molecular Physics, Kruislaan 407, 1098
 SJ Amsterdam, The Netherlands}
\date{\today}

\maketitle

\begin{abstract}

The surface free energy of ideal hard rods near curved hard surfaces is
determined to second order in curvature for surfaces of general
shape. In accordance with previous results for spherical and
cylindrical surfaces it is found that this quantity is non-analytical
when one of the principal curvatures changes signs. This prohibits
writing it in the common Helfrich form. It is shown that the
non-analytical terms are the same for any aspect ratio of the
rods. These results are used to find the equilibrium shape of vesicles
immersed in solutions of rod-like (colloidal) particles. The presence
of the particles induces a change in the equilibrium shape and to a shift
of the prolate-oblate transition in the vesicle phase diagram, which
are calculated within the framework of the spontaneous curvature
model. As a consequence of the special form of the energy contribution
due to the rods these changes cannot be accounted for by a simple
rescaling of the elastic constants of the vesicle as for solutions of spherical
colloids or polymers. 

\end{abstract}

\bigskip

\section{Introduction}

Recently Yaman et al. \cite{Yaman:97:PRL,Yaman:97} obtained the
surprising result that the surface tension $\gamma$ of a fluid of thin
hard rods at a spherical or cylindrical wall of radius $R$ is 
non-analytical at curvature $c=1/R=0$. They found different
expressions for $\gamma(c)$ for positive and negative curvatures. This
analysis that assumed ideal particles has been extended by Groh and
Dietrich \cite{rods} by taking into account the 
interactions between the particles. This leads to a substantial change of the curvature
dependence at medium and high particle densities, but a small singular contribution probably still
remains. In the present work we return to  the ideal limit but
generalize the problem in two respects. First, we explicitly study
arbitrarily shaped surfaces with locally varying curvatures. This is
especially important because due to the mentioned singularity  the curvature
dependence of the surface free energy cannot have the common Helfrich
form \cite{Helfrich:73}. Therefore it is not possible to derive the
general expression from the special cases of spherical and cylindrical
surfaces. Indeed, we find different analytical expressions depending
on the signs of the local curvatures which take on an unexpected form
especially at hyperbolic points. Second, we allow for a finite
thickness of the rod-like particles, with the known result for
spherical particles included as a limiting case.

One might ask if and how the curvature dependent surface free energy
is experimentally accessible. This is indeed the case for surfaces
that are free to adjust their curvature to the conditions provided by
the surrounding liquid. Such surfaces are given by vesicles which
consist of closed liquid bilayer membranes in an aqueous
solution. Their shapes are experimentally observable by phase-contrast
microscopy \cite{Doebereiner:97}. Theoretically vesicles are commonly
modeled as two-dimensional continuum surfaces with bending
elasticity. In the simplest model the elastic energy of the membrane
has just the Helfrich form. Different classes of equilibrium shapes
have been determined as local energy minima (for some especially
interesting examples see \cref{Jie:98}). For the most relevant
parameter ranges the corresponding phase diagram has been worked out
in detail \cite{Seifert:91}.

The presence of other particles in the solution gives rise to an
additional energy contribution. Due to a change of the effective
spontaneous curvature the vesicle may tend to bend towards or away
from the solute particles. The effects of absorbed and free
polymers and of spherical colloidal particles have been reviewed by
Lipowsky et al. \cite{Lipowsky:98}. Here we study the corresponding
problem for rod-like particles. In contrast to the other cases, here
the total elastic energy including the solute effects no longer has
the Helfrich form. Thus new equilibrium shapes arise and the phase
diagram is modified in a way that cannot be reduced to a simple
renormalization of the parameters.

\section{Surface free energy at arbitrarily shaped walls}

We study a system of hard spherocylinders of length $L$ and diameter
$D$ in the presence of a hard wall of general shape whose principal
radii of curvature $R_1$ and $R_2$ are large compared to $L$ and
$D$. The interactions between the rod-like particles are neglected,
i.e., we consider the dilute limit. In this section we calculate the
surface free energy of the rods, which is then used in the second part
of the paper to determine equilibrium shapes of vesicles immersed in
the rod solution.

Using the grand-canonical density-functional it can easily be shown
\cite{Yaman:97} that the surface contribution $\Omega_s$ to
the grand-canonical potential of the fluid is given by
\begin{equation} \label{Omsgen}
  \beta\Omega_s=\frac{\rho_b}{4\pi}\int d^3r d\omega \left(1-
  \frac{4\pi \hat\rho(\rv,\omega)}{\rho_b}\right),
\end{equation}
where $\hat\rho(\rv,\omega)$ denotes the density of rods whose center
of mass is at the point $\rv$ and whose orientation is
$\omega=(\theta,\phi)$ and $\rho_b$ is the particle density of the bulk
fluid far away from the wall. Due to the hard wall potential
$n(\rv,\omega):=4 \pi \hat\rho(\rv,\omega)/\rho_b$ has either the
value 0 or 1, depending on whether the configuration $(\rv,\omega)$ is
forbidden by the presence of the surface or not. The ideal
particles we assumed only ``feel'' the wall within a layer of
thickness $(L+D)/2$ so that the spatial integration in \eqref{Omsgen}
can be restricted to this layer. We introduce a local coordinate
system at each point of the surface whose $z$ axis points along the
normal direction into the rod solution and whose $x$ and $y$ axes are
aligned with the principal directions. As we are only
interested in the leading terms in the curvature we approximate the
position of the surface by
\begin{equation}
  z_s=-\half\left(\frac{x^2}{R_1}+\frac{y^2}{R_2}\right)
\end{equation}
where we introduced the convention that a curvature $c_i=1/R_i$ is
positive if the surface curves {\it away} from the fluid. Now
$\Omega_s$ can be written as a surface integral over the free energy
density $\omega_s$:
\begin{equation}
  \Omega_s=\int dS\,\omega_s(c_1,c_2)
\end{equation}
with
\begin{equation}
  \beta\omega_s(c_1,c_2)=\rho_b \int_0^{(L+D)/2} dz J(z,c_1,c_2)
  \frac{1}{4\pi}\int d\omega \left(1-n(z,c_1,c_2,\omega)\right).
\end{equation}
The Jacobian
\begin{equation}
  J(z,c_1,c_2)=\frac{(z+R_1)(z+R_2)}{R_1 R_2}=
  1+(c_1+c_2)z+c_1c_2z^2
\end{equation}
takes into account the change of the tangential area element $dS$ with
the normal distance $z$. The range of orientations $\omega$ for which
the particle  intersects the surface (i.e, for which
$n(z,c_1,c_2,\omega)=0$) for given curvatures $c_1$ and $c_2$ and
normal distance $z$ can be found by straightforward geometry.

\subsection{Infinitely thin rods}

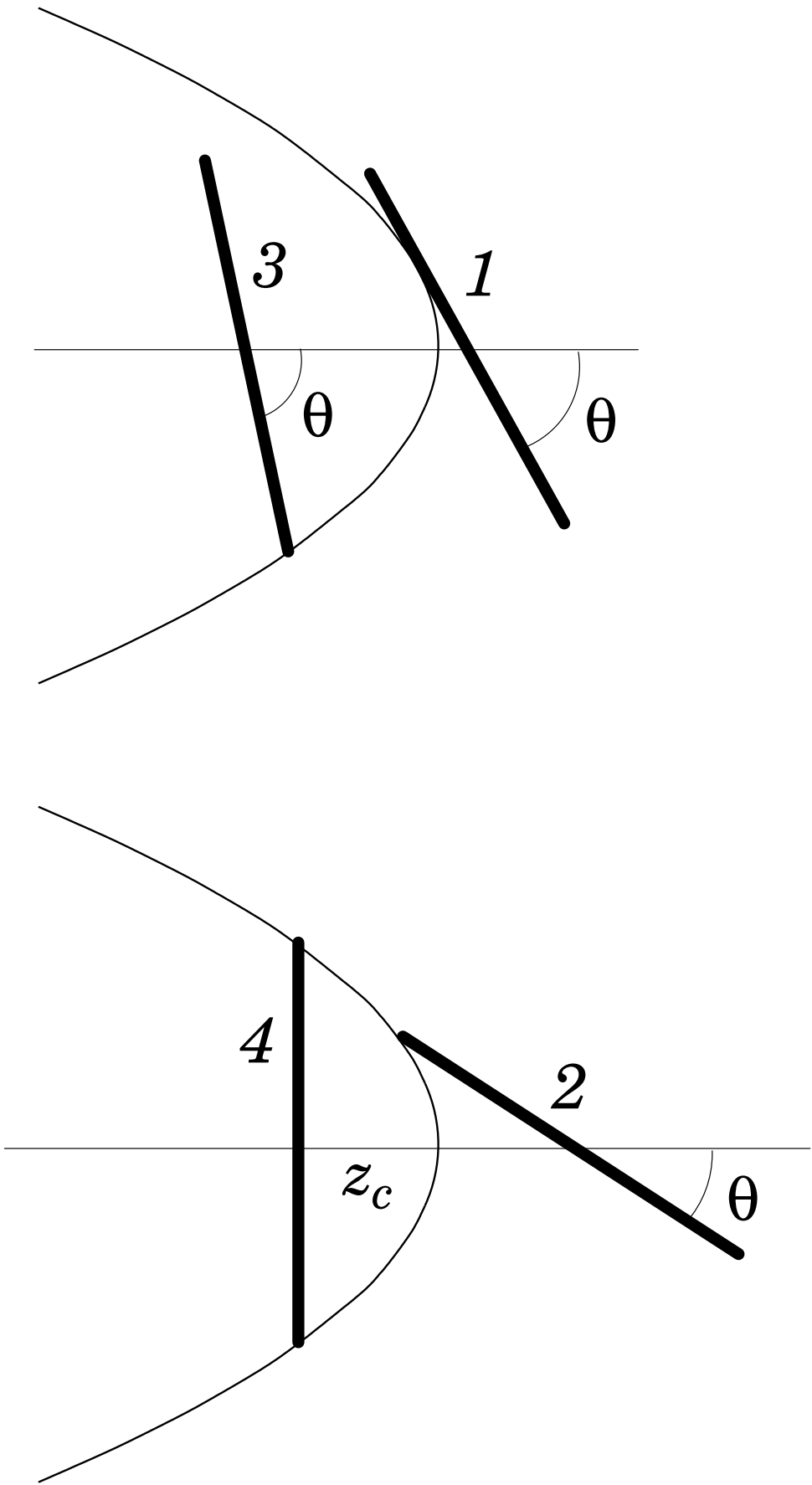
\begin{figure}
\caption{Geometries for the calculation of the available orientational
space of a rod near a general surface. For positive curvature the rod
touches the surface tangentially for small normal distances of its
center of mass (rod 1) and with its end at larger distances (rod
2). For negative curvature only end contact occurs (rod 3) and the rod
cannot come closer to the surface than the distance $z_c$ (rod 4). The
value of $x_{max}$ [Eqs.~(\pref{xmaxp}) and (\pref{xmaxm})] is the
cosine of the angle $\theta$ at contact.}
\label{fig:touch}
\end{figure}

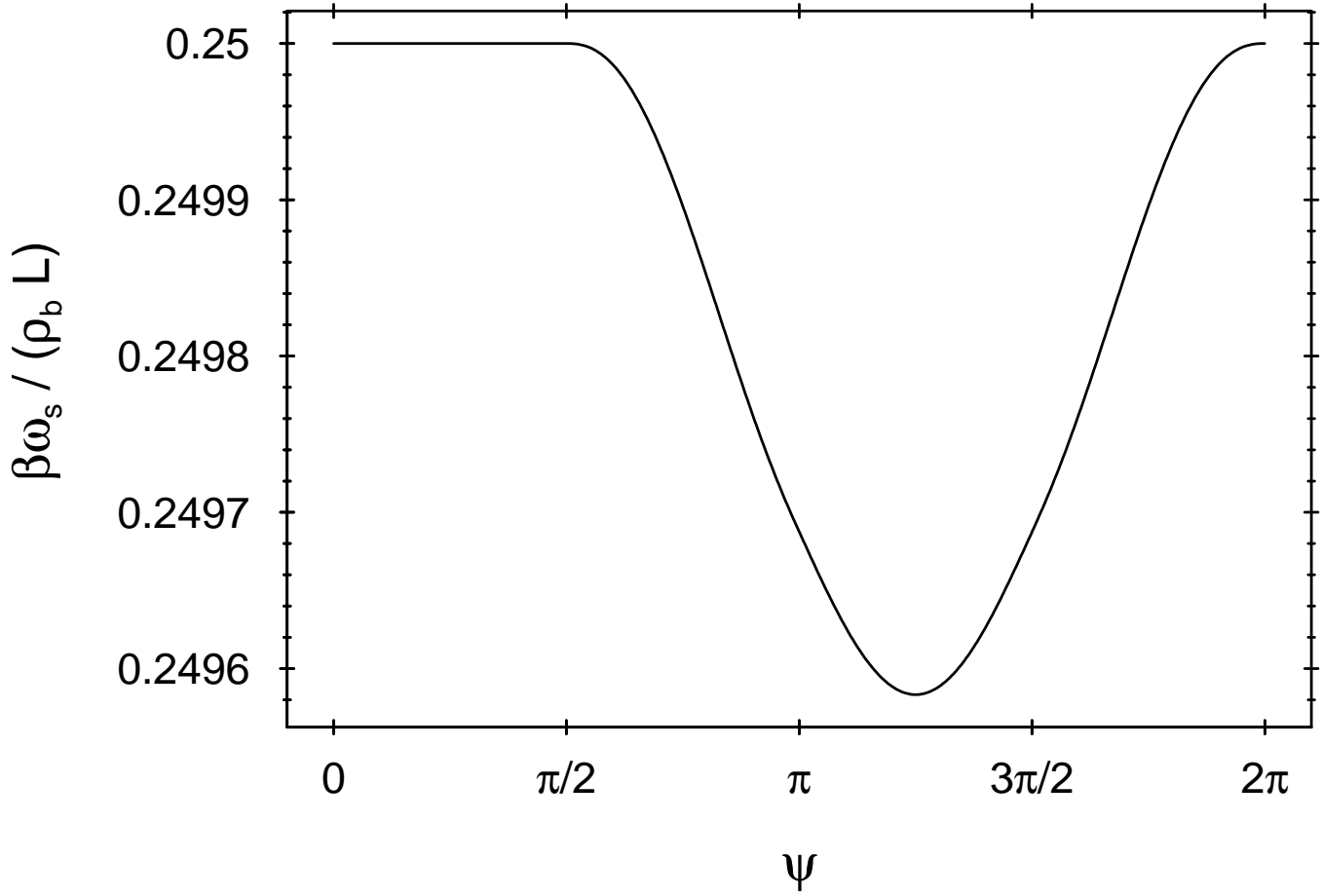
\begin{figure}
\caption{Curvature dependence of the surface free energy density
$\omega_s(c_1,c_2)$ for $c_1=c\cos\psi$ and $c_2=c\sin\psi$. The value
of $c$ is kept fixed ($cL=0.2$) and $\psi$ is varied from $0$ to
$2\pi$ providing a continuous path through the four possible sign
combinations of $c_1$ and $c_2$ for which four different analytical
expressions [Eqs.~(\pref{fspp})--(\pref{fspm})] apply. The curve is
non-analytic at $\psi=\frac{\pi}{2},\pi,\frac{3\pi}{2},2\pi$. The
third derivative with respect to $\psi$ diverges on one side of these points.}
\label{fig:fspsi}
\end{figure}

We first consider the limit of infinitely thin rods ($D=0$). At given
azimuthal angle $\phi$ around the normal direction we determine the
maximum allowed value $x_{max}$ for the cosine of the polar angle, at
which the rod just touches the surface. This is a two-dimensional
problem within the plane $\phi={\rm const}$ that intersects the
surface in the parabola $z_s=-c \rho^2$ ($\rho$ is the tangential
coordinate within the plane) with
\begin{equation}
  c=c(\phi)=c_1\sin^2\phi +c_2\cos^2\phi
\end{equation}
Note that $c(\phi)$ can be positive or
negative depending on the signs of $c_1$ and $c_2$. Figure~\ref{fig:touch} shows that for positive curvatures
there are two possibilities how the thin rod can touch the surface:
tangentially for $z<z_c$ and  with its end for $z>z_c$. It is easy to
show that
\begin{equation} \label{xmaxp}
  x_{max}=\left\{
   \begin{array}{ll}
     (1+\frac{1}{2 c z})^{-1/2}= \sqrt{2 c z}+O(c^{3/2}), & z<z_c
   \\
   -\frac{2}{c L}+\sqrt{1+\frac{8 z}{cL^2}+\frac{4}{c^2 L^2}}
   = \frac{2 z}{L}+(\frac{L}{4}-\frac{z^2}{L})
   c+(\frac{z^3}{L}-\frac{z L}{4}) c^2+O(c^3)
   , & z>z_c
   \end{array}\right.
\end{equation}
with
\begin{equation}
  z_c=\frac{1}{4 c}(-1+\sqrt{1+c^2 L^2})= \frac{c L^2}{8}+O(c^3)
\end{equation}
For negative curvatures the rod cannot approach the surface closer
than $z_c=-c L^2/8$ so that $x_{max}=0$ for $z<z_c$ while for $z>z_c$ one
has
\begin{equation} \label{xmaxm}
  x_{max}=\frac{2z}{ L}+(\frac{L}{4}-\frac{z^2}{L})
  c+(\frac{z^3}{L}-\frac{z L}{4}) c^2+O(c^3)
\end{equation}
as before. Note that $x_{max}$ is analytic at $c=0$ only if $z>z_c$
while different expressions apply for $c\lessgtr 0$ if $z<z_c$. The
surface free energy density now follows by integration over $\phi$:
\begin{equation}
  \frac{\beta\omega_s}{\rho_b}(c_1,c_2)=\frac{1}{2\pi}\int_0^{2\pi} d\phi
  \int_0^{L/2} dz J(z,c_1,c_2) (1-x_{max}(z,c(\phi))).
\end{equation}
Up to terms quadratic in the curvature one obtains
\begin{equation} \label{fspp}
  \frac{\beta\omega_s}{\rho_b}=\frac{L}{4}
   \qquad\mbox{for } c_1,c_2>0
\end{equation}
and
\begin{equation} \label{fsmm}
  \frac{\beta\omega_s}{\rho_b}=\frac{L}{4}-\frac{L^3}{128}((c_1+c_2)^2-\frac{4}{3} c_1 c_2)
   \qquad\mbox{for } c_1,c_2<0
\end{equation}
If $c_1>0$ and $c_2<0$ the function $c(\phi)$ changes sign at the
angle $\phi_0=\arctan\sqrt{-c_2/c_1}$ and the $\phi$ integration must
be split accordingly. After some algebra one finds
\begin{eqnarray} \label{fspm}
  \frac{\beta\omega_s}{\rho_b}&=&\frac{L}{4}-\frac{L^3}{192\pi}\left[
  3\sqrt{-c_1 c_2}(c_1+c_2)+(3c_1^2+2 c_1
  c_2+3c_2^2)\arctan\sqrt{-c_2/c_1} \right] \\
  & & 
  \qquad\mbox{for } c_1>0, c_2<0. \nn
\end{eqnarray}
Of course the same formula with $c_1$ and $c_2$ interchanged applies
for $c_1<0$ and $c_2>0$. Thus depending on the signs of the principal
curvatures $\omega_s$ is given by \eqref{fspp}, \klref{fsmm}, or
\klref{fspm}. Equation~\klref{fsmm} was already derived by Yaman et
al. (see Eq.~(A.9) in \cref{Yaman:97}) who also showed that
\eqref{fspp} is exact for a convex surface to all orders in
the curvature. Our new result \eqref{fspm} provides a continuous
connection between \eqref{fspp} and \eqref{fsmm}. This is demonstrated
in \abbref{fig:fspsi} where $\omega_s$ is plotted for $c_1=c \cos\psi$
and $c_2=c \sin\psi$ as a function of $\psi$ for fixed $c$. Obviously
the presence of the rods favors surfaces with two negative principal
curvatures. Terms linear in the curvature are absent in all cases. The
quadratic terms are non-analytic if one of the $c_i$'s changes sign,
which, inter alia, precludes writing the curvature dependence of the surface free energy in the
common Helfrich form
\begin{equation} \label{FHel}
  F_{Hel}=\int dS (\half \kappa (c_1+c_2-C_0)^2+\bar\kappa c_1 c_2)
\end{equation}
with the bending rigidities $\kappa$ and $\bar\kappa$ and the
spontaneous curvature $C_0$. If one restricts oneself to equal signs
of $c_1$ and $c_2$ we have rigorously shown here that the Helfrich
form can be applied with $\beta\kappa/\rho_b=-L^3/64$ and
$\beta\bar\kappa/\rho_b=L^3/96$ for negative and, trivially,
$\kappa=\bar\kappa=0$ for positive curvatures. However, surfaces of
vesicles typically exhibit regions with differing signs, so that a
global mapping to the Helfrich form is no longer possible.

\subsection{Rods of finite thickness} \label{finiteD}

The case of spherocylindrical particles with finite thickness $D$ can
be reduced to the problem for $D=0$ discussed above if one realizes
that the surface of the spherocylinder consists of all points that
have a distance $D/2$ from the line segment of length $L$ which
connects the centers of the hemispherical caps. Thus if the
spherocylinder touches the surface $S$ this line segment touches a
parallel surface $S'$ shifted by $D/2$. The curvature radii of $S'$
and $S$ are simply related by $R_i'=R_i+D/2$, i.e., $c_i'=c_i/(1+c_i
D/2)$ \cite{DoCarmo:76}, so that for $z>D/2$
\begin{equation}
  n(z,c_1,c_2,D)=n(z-D/2,c_1/(1+c_1 D/2),c_2/(1+c_2 D/2),0).
\end{equation}
The particles cannot approach the wall closer than $z=D/2$. The
inaccessible range yields the $L$ independent contribution
\begin{equation}
  \frac{\beta\omega_s^{(1)}}{\rho_b}=\int_0^{D/2} dz\,J(z,c_1,c_2)
   =\frac{D}{2}+\frac{D^2}{8}(c_1+c_2)+\frac{D^3}{24} c_1 c_2
\end{equation}
to the free energy density. By using the fact that
\begin{equation}
  J(z'+D/2,c_1,c_2)=J(z',c_1',c_2')(1+c_1 D/2)(1+c_2 D/2)
\end{equation}
 one
obtains for the remaining $z$ integral from $D/2$ to $(L+D)/2$
\begin{equation}
  \omega_s^{(2)}(c_1,c_2,D)= \left(1+\frac{c_1
  D}{2}\right)\left(1+\frac{c_2 D}{2}\right) \omega_s(c_1',c_2',0).
\end{equation}
Now the above results for infinitely thin rods can be employed to give
the final result
\begin{equation}
  \frac{\beta\omega_s}{\rho_b}=\frac{D}{2}+\frac{L}{4}+(\frac{D^2}{8}+\frac{DL}{8})(c_1+c_2)+(\frac{D^3}{24}+\frac{D^2 L}{16}) c_1 c_2 +\Delta f_s
\end{equation}
with
\begin{eqnarray}
  \Delta f_s & = & 0, \qquad c_1,c_2>0 \\
  \Delta f_s & = & -\frac{L^3}{128}((c_1+c_2)^2-\frac{4}{3} c_1 c_2),
  \qquad c_1,c_2<0 \\
  \Delta f_s & = & -\frac{L^3}{192\pi}\left[
  3\sqrt{-c_1 c_2}(c_1+c_2)+(3c_1^2+2 c_1
  c_2+3c_2^2)\arctan\sqrt{-\frac{c_2}{c_1}} \right], \qquad
  c_1>0, c_2<0.
\end{eqnarray}
It is remarkable that all the additional terms which contain powers of
$D$ are analytic in $c_1$ and $c_2$. The singular contributions $\Delta
f_s$ are exactly the same as in the limit $D\to0$. In this sense no
qualitatively new features arise when the thickness of the rods is
taken into account. Therefore we only consider thin rods in the
second part of this paper. The special case of hard non-interacting
spheres is also contained in the above analysis for $L=0$. Only in
this case the full $\omega_s$ is analytic and thus lends itself to the
Helfrich expansion. Our result for the spheres is in agreement with
previous calculations \cite{Yaman:97:PRL,Eisenriegler:96}.

\section{Vesicle shapes in a rod solution}

We now consider a vesicle immmersed in a solution of rodlike
colloids. It is assumed that the rods are not absorbed on the membrane
and that their interaction can be approximated by the hard wall
model. The bending energy of the vesicle consists of two
contributions: First, the internal bending rigidity of the membrane
which is described in the so-called spontaneous curvature model as
given by \eqref{FHel} (For other possibilities see, e.g.,
\cite{Doebereiner:97,Miao:94}). Second, the presence of the rodlike
particles gives rise to the surface free energy determined in the
previous section.  If the vesicle is immersed in a solution of
polymers \cite{Hanke:98} or diluted spherical colloids this second
contribution will have the same analytical form as the first and
therefore lead to a renormalization of the rigidity coefficients and
the spontaneous curvature of the ``free'' vesicle. Then the equilibrium
shapes are those determined for the pure spontaneous curvature model
\cite{Seifert:91} and the addition of the solute only shifts the
system to a different point in the phase diagram. However, in the
present case the energy contribution due to the rods cannot be written
in the Helfrich form so that new equilibrium shapes will arise.  As
shown in Sec.~\ref{finiteD} a non-zero thickness of the rods can also
be taken into account by using renormalized coefficients in
$F_{Hel}$. Therefore in the following we only use the expressions for
infinitely thin rods.

Because the energy scale associated with the bending rigidity $\kappa$
is much smaller than the energies necessary to change significantly
the area $A$ or the volume $V$ of the vesicles, these quantities are
regarded as fixed. Thus the equilibrium shape is determined by
minimization of $F_{tot}=F_{Hel}+\Omega_s$ under the constraint of
fixed $A$ and $V$. Due to the scale invariance of $F_{tot}$ the
equilibrium shape depends only on the dimensionless parameters
\begin{equation} \label{defvc0x}
  v=\sqrt{36\pi}\frac{V}{A^{3/2}}, \qquad
  c_0=C_0 \sqrt{\frac{A}{4\pi}}, \qquad
  x=\frac{\rho_b L^3}{\beta\kappa}.
\end{equation}
A sphere is characterized by $v=1$, while for any other shape
$v<1$. The parameter $x$ measures the relative importance of the
contribution $\Omega_s$ due to the rods. The Gaussian curvature term
proportional to $\bar\kappa$ in \eqref{FHel} is constant for
topologically equivalent shapes and will be omitted 
since we consider only shapes with the same topology as the sphere. We
will focus on the region $v\lesssim 1$ and moderate values of $c_0$
where the equilibrium shapes of the  pure spontaneous curvature model
are axisymmetric prolates or oblates with up-down symmetry, separated by a first-order transition
\cite{Seifert:91}. 

Rather than deriving and solving the exact
Euler-Lagrange shape equations we parametrize the surfaces by simple
two-parameter functions. In cylindrical coordinates $(r,z,\phi)$
around the axis of symmetry we write \cite{Helal:98}
\begin{eqnarray}
  r(z)=a \sqrt{1-z^2}\sqrt{1+bz^2} & \qquad & \mbox{prolates} \\
  z(r)=a \sqrt{1-r^2}\sqrt{1+br^2} & \qquad & \mbox{oblates}
\end{eqnarray}
These shapes reduce to ellipsoids for $b=0$ while increasing the value of $b$
leads to a  bulging which finally yields non-convex shapes for
$b>1$. The parameter
$a$ is the aspect ratio of the shape, i.e., the ratio of its diameter
in the equatorial plane and its length along the symmetry axis (or
vice versa for oblates). We emphasize  that the bending energy
does not depend on the size of the vesicles so that the length of one
axis can be arbitrarily set to unity. The corresponding vesicle
volumes are
\begin{eqnarray}
  V_{pro} & = & \frac{4\pi}{3} a^2 (1+\frac{b}{5}) \\
  V_{obl} & = & \frac{\pi a}{2 b} \left[b-1+\frac{(b+1)^2}{2\sqrt{b}}
  \left(\frac{\pi}{2}+\arcsin\frac{b-1}{b+1}\right)\right]
\end{eqnarray}
and the areas
\begin{eqnarray}
  A_{pro} & = & 4\pi a\int_0^1 dz\sqrt{(1-z^2)(1+b z^2)+a^2 z^2
  (1-b+2bz^2)^2} \\
  A_{obl} & = & 4\pi \int_0^1 dr\,r\sqrt{1+a^2 r^2 \frac{(1-b+2b
  r^2)^2}{(1-r^2)(1+b r^2)}}.
\end{eqnarray}
 The
principal curvatures are
\begin{equation}
  c_1^{pro}=r(z) \sqrt{1+r'(z)^2},\qquad
  c_2^{pro}=-\frac{(1+r'(z)^2)^{3/2}}{r''(z)}
\end{equation}
for prolates and
\begin{equation}
  c_1^{obl}=-\frac{r\sqrt{1+z'(r)^2}}{z'(r)},\qquad
  c_2^{obl}=-\frac{(1+z'(r)^2)^{3/2}}{z''(r)}
\end{equation}
for oblates.
All curvatures are positive for $b<1$. In the prolate case $c^{pro}_2$
becomes negative for $z<z_c$ where $z_c$ is the positive solution of
\begin{equation} \label{zc}
  -1+b^2 z^4 (-3+2 z^2)+b(1-6 z^2+3 z^4)=0.
\end{equation}
For oblates three ranges must be distinguished: for $r<r_{c1}$, where
$r_{c1}$ is again determined by \eqref{zc}, both curvatures are
negative. Between $r_{c1}$ and $r_{c2}=\sqrt{(b-1)/(2b)}$ $c^{obl}_2$
is positive and $c^{obl}_1$ is negative, while for $r>r_{c2}$ both
curvatures are positive. The rod contribution to the elastic energy is
computed from
\begin{equation}
  \Omega_s=\int dS\,\omega_s(c_1,c_2)
  =\left\{ \begin{array}{ll}
  4\pi \int_0^1 dz\,r(z) \sqrt{1+r'(z)^2}\,
  \omega_s(c_1^{pro}(z),c_2^{pro}(z)) & \mbox{prolates} \\
  4\pi \int_0^1 dr\,r\sqrt{1+z'(r)^2}\,
  \omega_s(c_1^{obl}(r),c_2^{obl}(r)) & \mbox{oblates}
 	   \end{array} \right.
\end{equation}
The equilibrium shapes follow by numerical minimization of the total
elastic energy $F_{tot}(a(b,v),b)$ with respect to $b$ where $a(b,v)$
is obtained by numerical solution of the first equation of
\eqref{defvc0x}. 

If the rods are inside instead of outside the vesicle the signs of the
curvatures $c_i$ have to be reversed in the computation of
$\Omega_s$. For the case of particles on {\it both} sides of the
membrane $\omega_s(c_1,c_2)$ must be replaced by
$\omega_s(c_1,c_2)+\omega_s(-c_1,-c_2)$ which, interestingly, has the
form given by \eqref{fsmm} for {\it all} signs of the $c_i$. Hence
here the Helfrich form is valid, as has been surmised by Yaman et
al. \cite{Yaman:97:PRL}, so that for $c_0=0$ the equilibrium shape
does not depend on the rod concentration $x$.

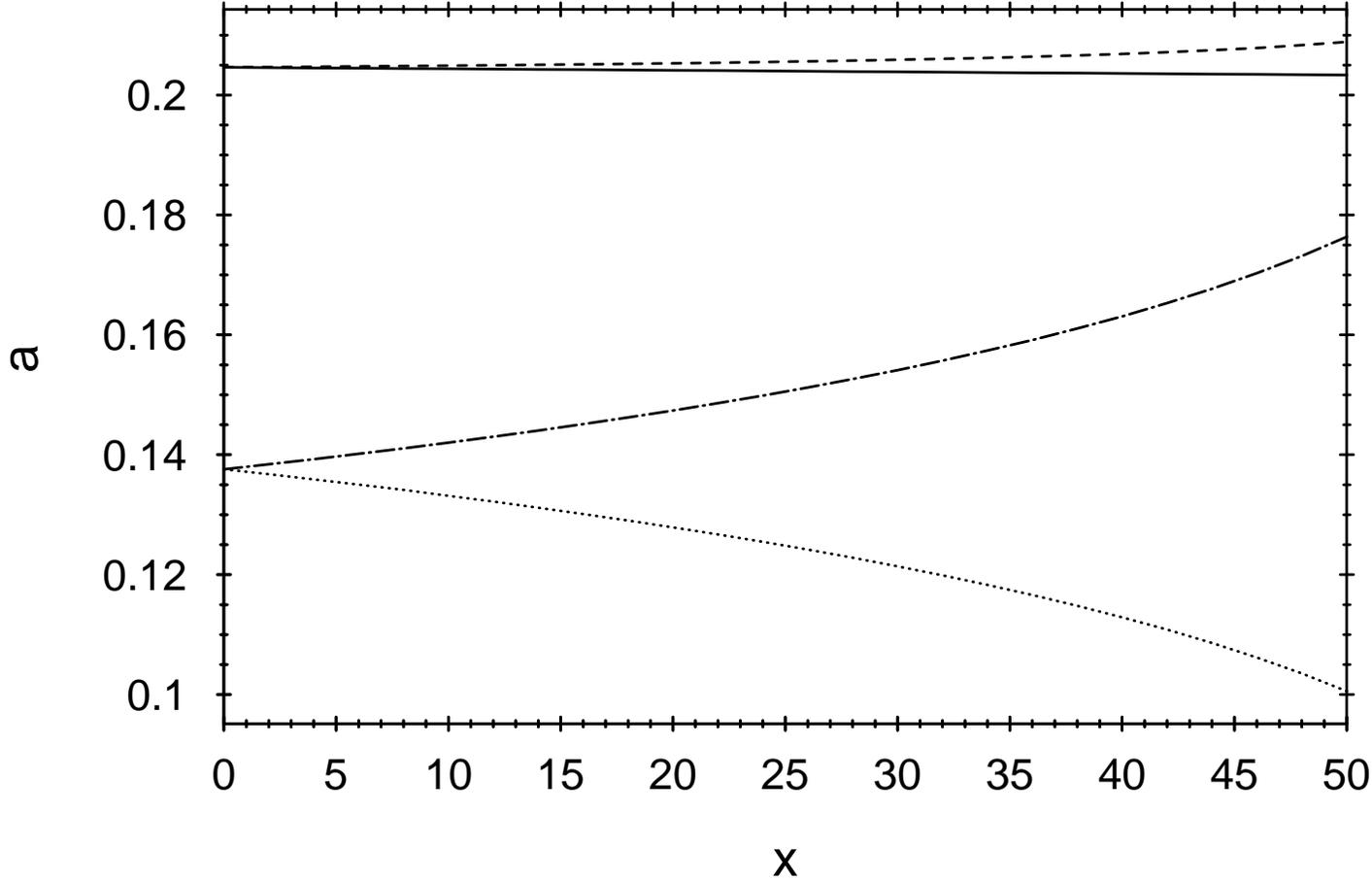
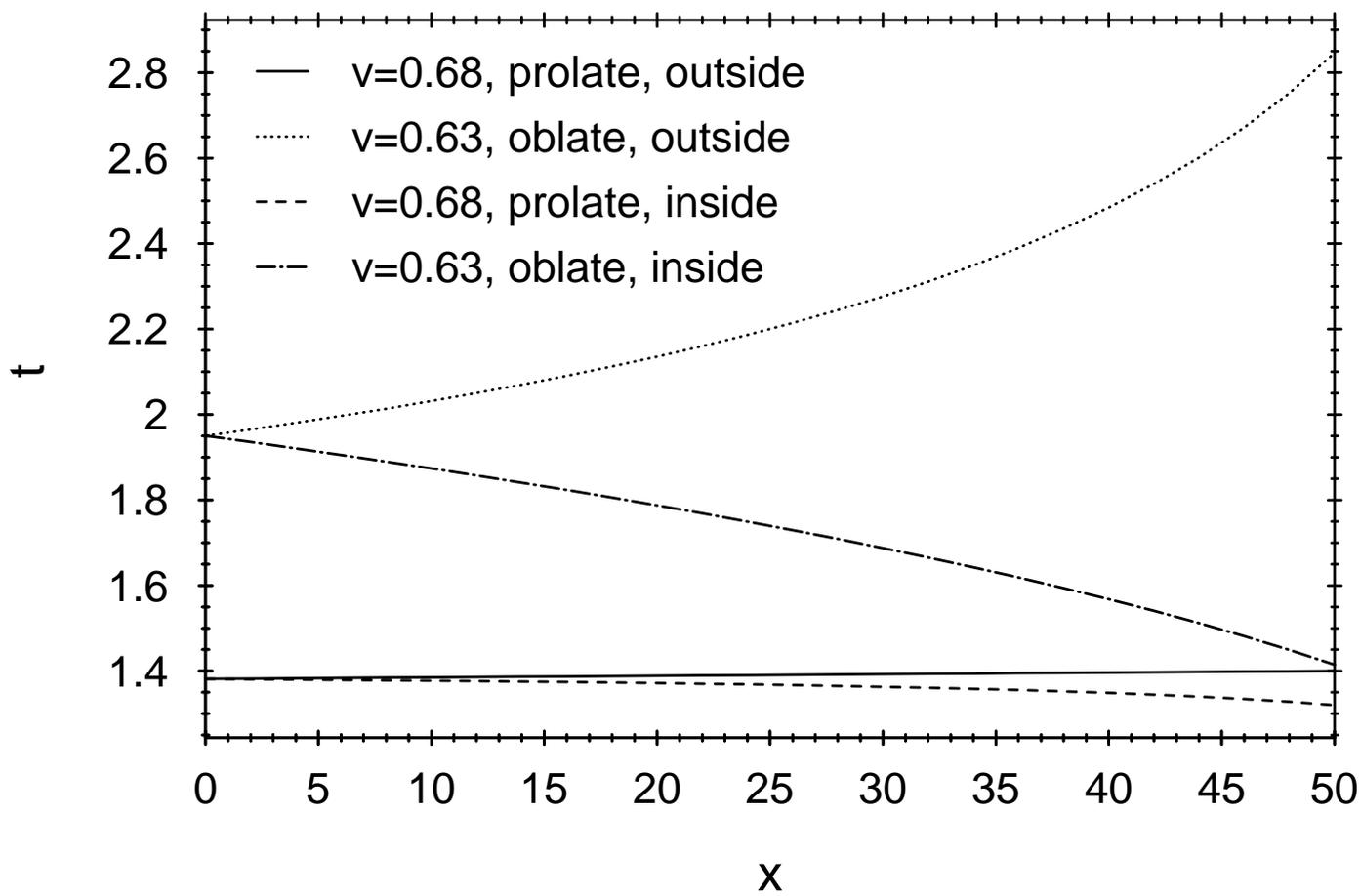
\begin{figure}
\caption{Aspect ratio $a$ and bulging parameter $t$ (see main text) of
the equilibrium shape as a function of the reduced concentration $x$
of rods inside and outside the vesicle for prolates and oblates at two
reduced volumina $v$ ($c_0=0$) close to the transition point.}
\label{fig:shape}
\end{figure}

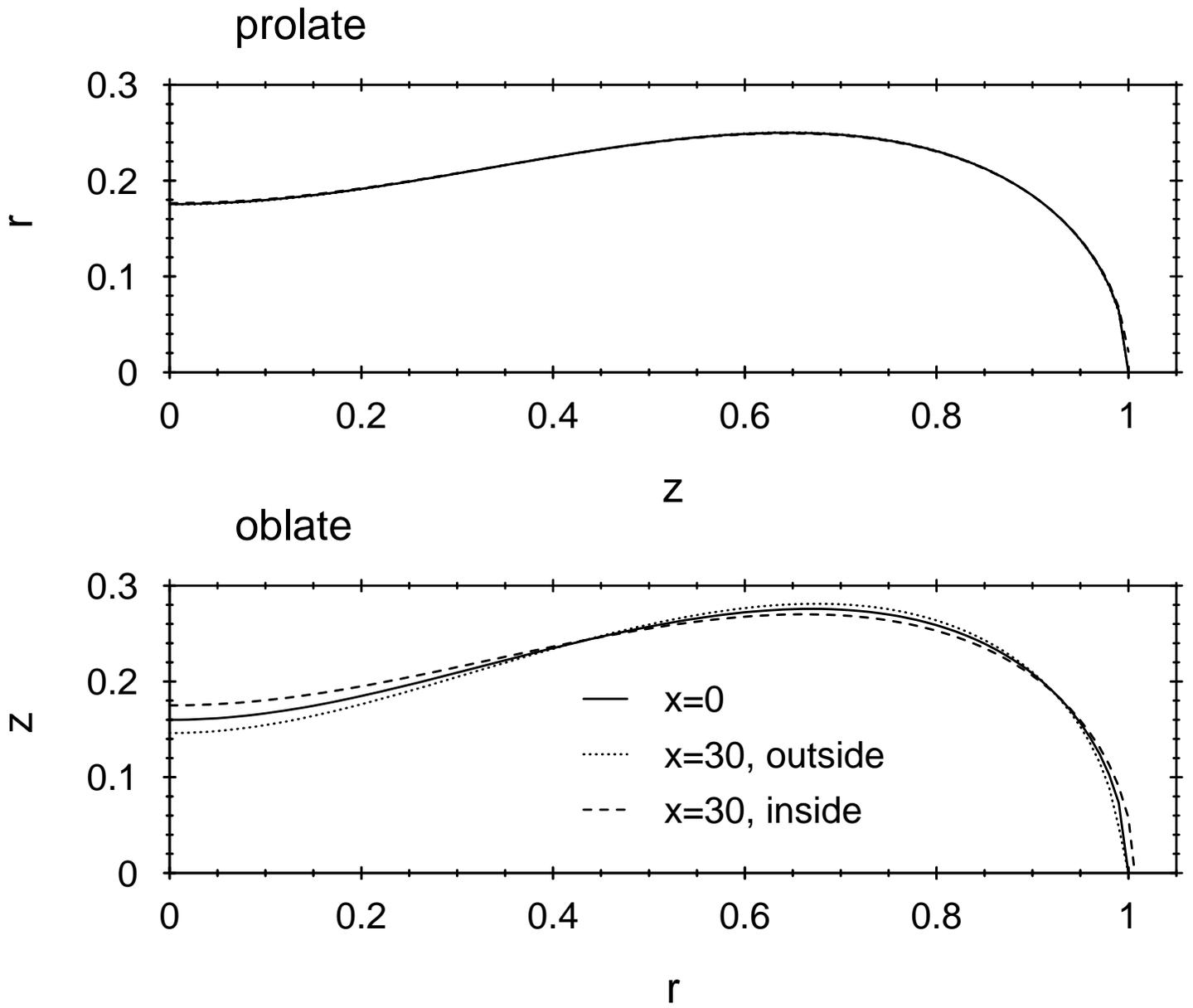
\begin{figure}
\caption{Equilibrium shapes for $c_0=0$ at $v=0.648$ which corresponds
to the transition point if no rods are present. Only one quarter of
the contours is drawn, the remaining parts follow by symmetry. Note
that the
symmetry axis is drawn horizontally for the prolates and vertically
for the oblates. Within each part of the figure the shapes are scaled
to the same volume and area. The presence of rods outside or inside
the vesicle induces a modification of the shape, which, however, for
the prolates is hardly visible on this scale.}
\label{fig:shpcx}
\end{figure}

We first discuss the results for zero spontaneous curvature. If $b>1$
the shape can be characterized by a ``bulging'' parameter $t=(b+1)/2
b^{1/2}$ which is equal to the ratio of the maximum of $r(z)$ [$z(r)$]
and its value in the equatorial plane [along the symmetry axis] for
prolates [oblates]. In \abbref{fig:shape} the quantities $a$ and $t$ are shown as a
function of $x$ for a prolate and an oblate at $v$ values in the
vicinity of the phase transition. Rods outside a prolate tend to
decrease $a$ and increase $t$ thereby narrowing the waist of the
vesicle. Similarly an oblate develops stronger ``dips'' at the
symmetry axis in order to increase the range and degree of negative
curvature. These results confirm our previous observation that the
surface prefers to bend towards the rods. Figure~\ref{fig:shpcx}
shows some examples for shapes without and with rods outside and
inside the vesicle. The changes of $a$ and $t$ are much smaller for
prolates (0.4--1.3\% at $x=30$) than for oblates (10--15\%).

Identifying the most probable shape with the lowest energy shape is
problematic when the energy is a non-analytic function of the shape
which might be non-quadratic for small deviations from the energy minimum
\cite{Hanke:priv}. However, numerically the function
$\Omega_s(a(b,v),b)$ shows no traces of non-analyticity. In fact, the
integrand $\omega_s(c_1,c_2)$ is singular only on a set of measure
zero (on the curves determined by $c_1c_2=0$) so that the singularity is
probably removed by the integration.

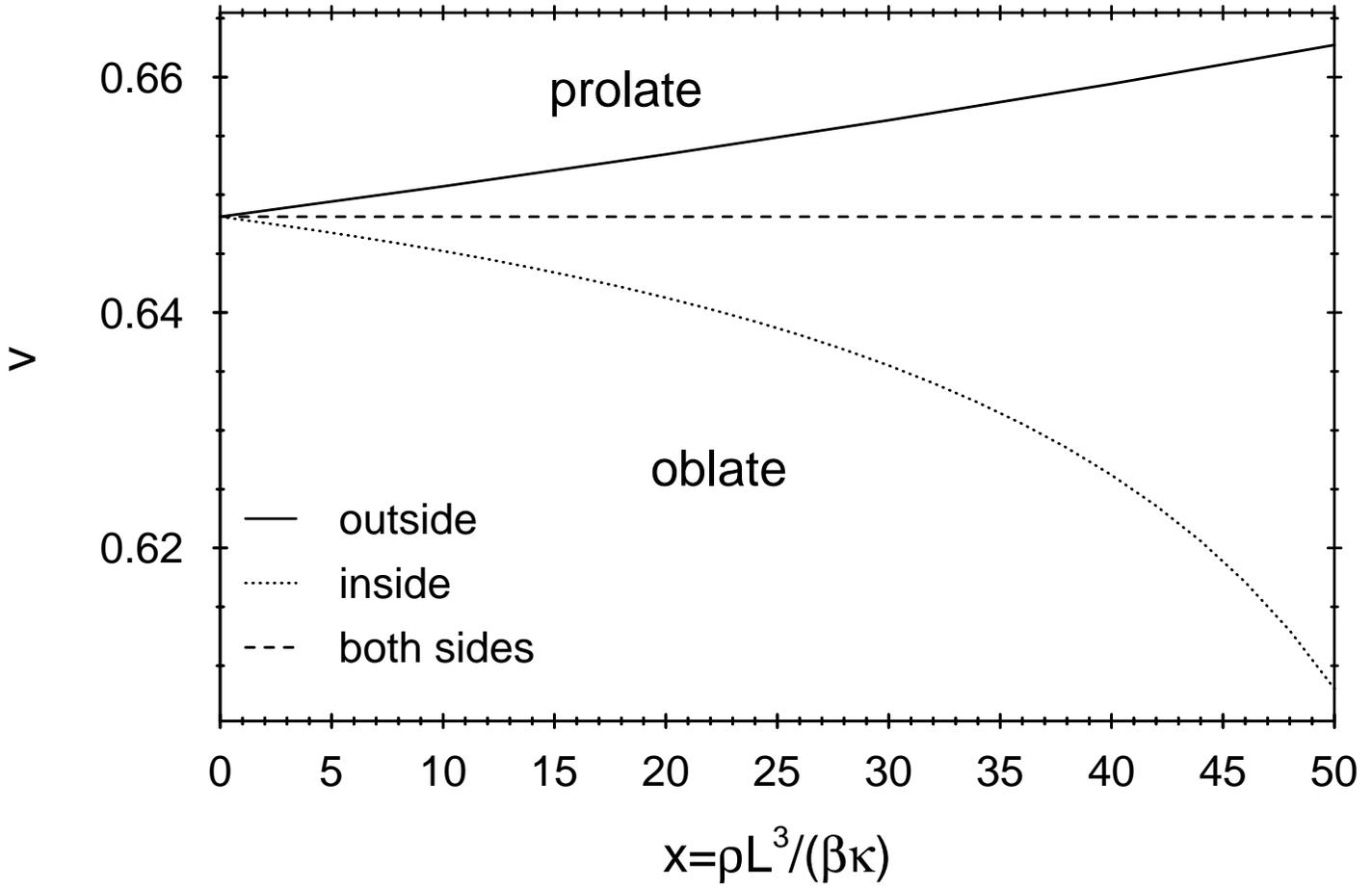
\begin{figure}
\caption{Phase boundary between prolates and oblates for $c_0=0$ as a
function of the dimensionless rod concentration $x$ outside, inside,
or on both sides of the vesicle. Particles outside favor oblate
shapes, particles inside prolate shapes. If particles are present on
both sides their contribution to the elastic energy has the Helfrich
form, too, so that the equilibrium shapes and the boundary are not altered.}
\label{fig:phas}
\end{figure}

Without rods the phase transition between prolates and oblates takes
place at $v=0.648$ within our approximation which is very close to the
value $v=0.651$ obtained by an exact minimization
\cite{Seifert:91}. The coexisting shapes (see \abbref{fig:shpcx}) both
are strongly non-spherical with an aspect ratio $a\simeq0.17$ which
indicates a pronounced first-order character of the transition. As
shown in \abbref{fig:phas} the rods shift the transition to larger $v$
if they are outside the vesicle and to smaller $v$ inside. As
explained above rods on both sides effectively just change the value
of $\kappa$ but do not influence the phase diagram.

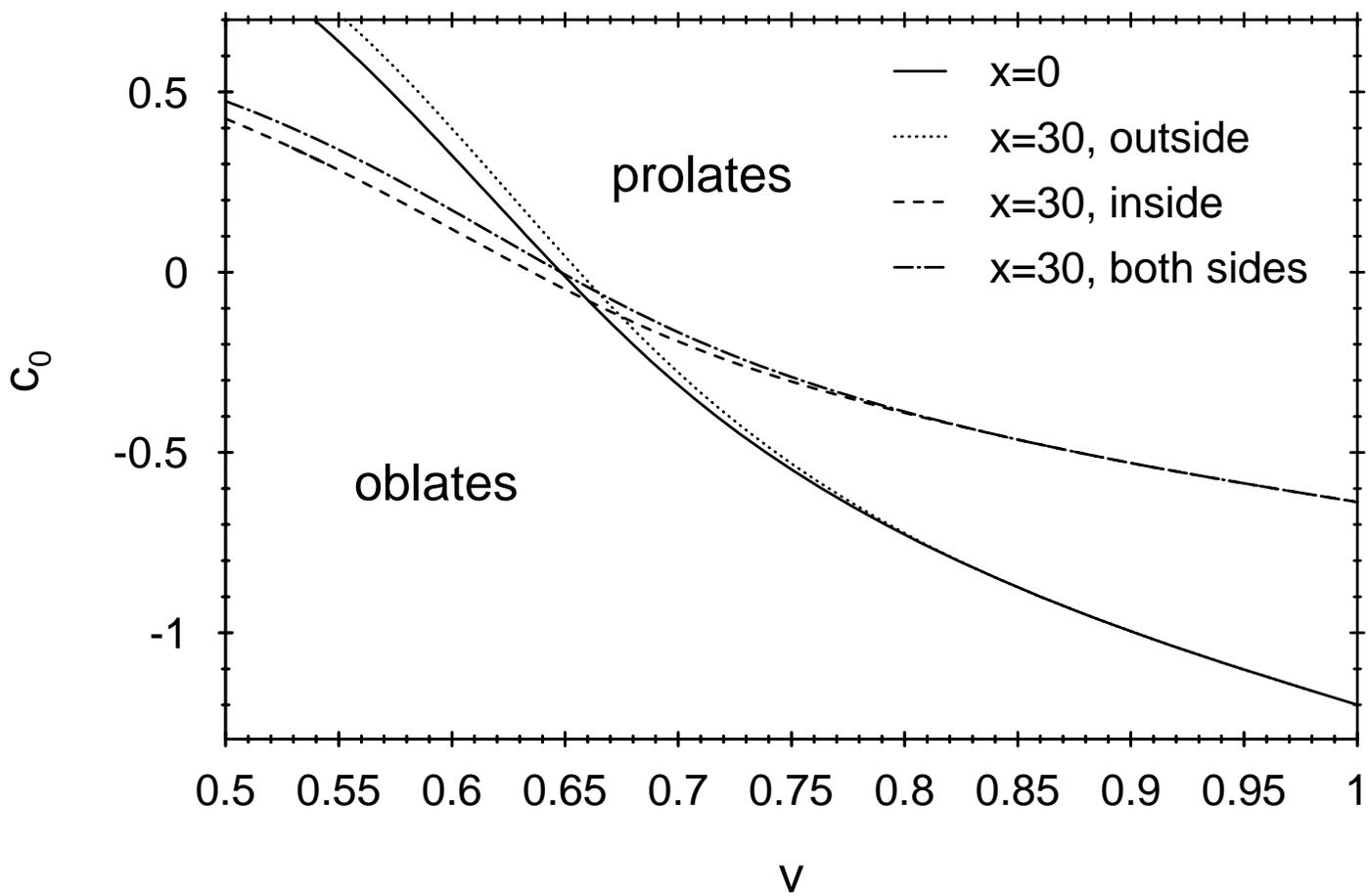
\begin{figure}
\caption{Phase diagram in the $v,c_0$ plane for vesicles in the
absence and presence of rods. The lines approach the maximum value
$v=1$ at $c_0=-6/5$ for $x=0$ or rods outside, but at
$c_0=-6/5(1-x/64)$ for rods inside or on both sides (see main text).}
\label{fig:phasc0}
\end{figure}

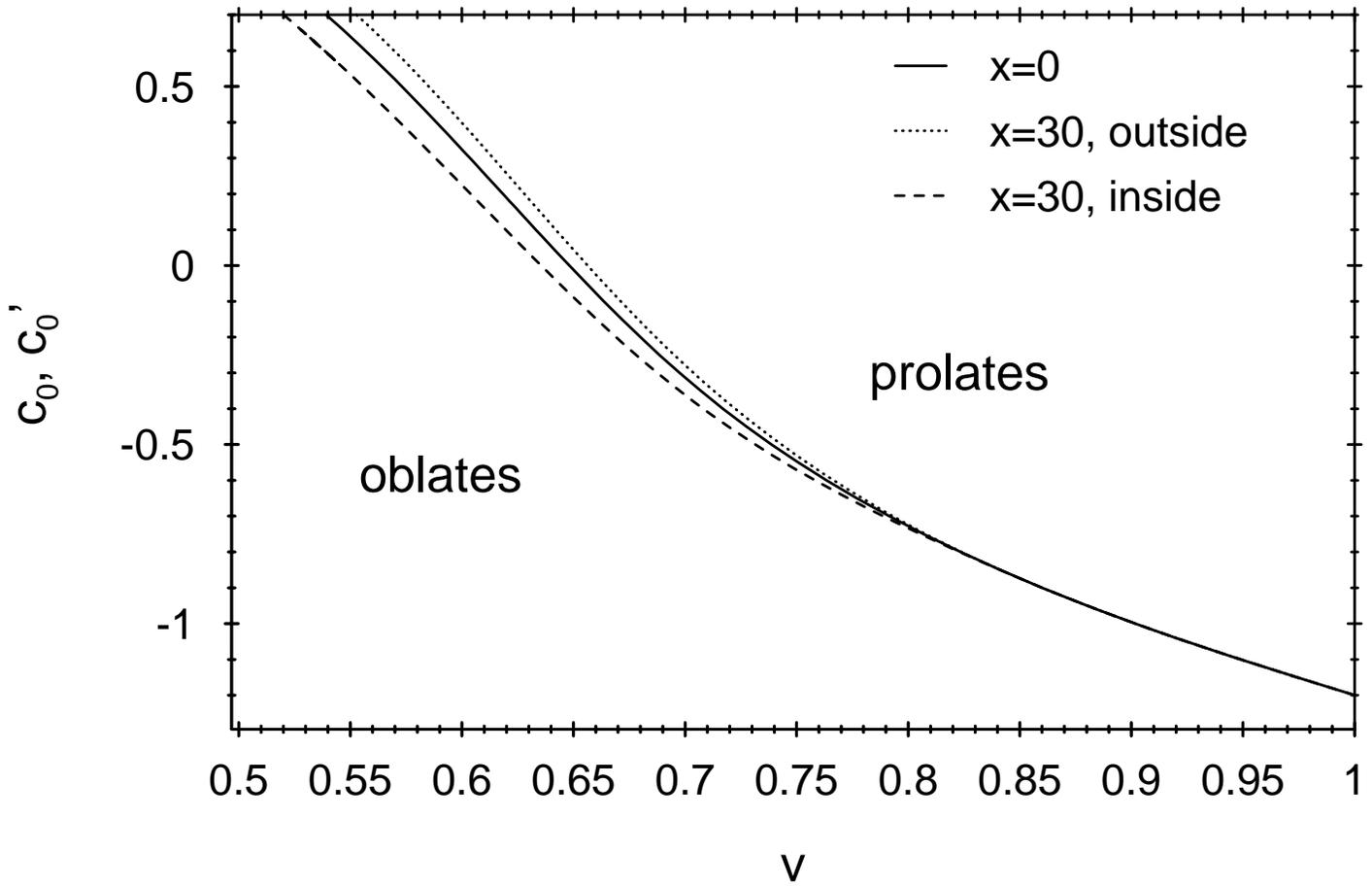
\begin{figure}
\caption{The same phase diagram as in \abbpref{fig:phasc0} but plotted
in terms of the effective spontaneous curvature $c'_0$ (\eqpref{c0s})
for rods inside and on both sides of the vesicle. The curve for the
latter case coincides with that for the free vesicle ($x=0$) in this
representation.}
\label{fig:phasc0s}
\end{figure}

We now turn to the more general case of non-zero spontaneous curvature
$c_0$. Negative values of $c_0$ favor oblate shapes so that the
transition point moves to higher volume to surface ratios $v$
(\abbref{fig:phasc0}). It has been shown rigorously that the phase
boundary approaches $v=1$ at $c_0=-6/5$
\cite{Milner:87,Seifert:91}. By a series expansion of the elastic
energy around $v=1$ one finds that this also hold within the present
parametrization. The presence of rods outside the vesicle shifts the
transition line to larger $v$. However, no shift occurs for $v>0.87$
because in this range both coexisting shapes are convex ($b<1$) so
that $\Omega_s=\mbox{const}$. Rods on both side of the membrane give
rise to an additional Helfrich like term with
$\beta\kappa_{rod}=-\rho_b L^3/64$ and $c_{0,rod}=0$. Therefore the
effective bending rigidity is $\kappa'=\kappa+\kappa_{rod}$ and the
effective spontaneous curvature is 
\begin{equation} \label{c0s}
 c'_0=\frac{c_0}{1-x/64}.
\end{equation}
For this reason the curve for particles on both sides differs from the
curve for $x=0$ in \abbref{fig:phasc0} (except at $c_0=0$) and reaches
$v=1$ at $c_0=-6/5 (1-x/64)$, whereas both curves are identical if
plotted as a function of $c'_0$ instead of $c_0$
(\abbref{fig:phasc0s}). We note that $\kappa'$ becomes negative for
$x>64$. In this case the present analysis breaks down and higher order
terms in the curvature must be taken into account. If the rods are
restricted to the inside of the vesicle the same effective Helfrich
form applies as long as only convex shapes are considered. This
explains why the corresponding phase boundary is equal to that for
rods on both sides at large $v$. However, for smaller $v$ shapes with
differing signs of the curvatures occur which cannot be
accounted for by a simple rescaling of the Helfrich coefficients and
which shift the transition line in \abbref{fig:phasc0s} to lower $v$
compared to the pure Helfrich case.

\section{Discussion}

In summary, we have shown that if rod-like particles are present on
the outer or inner side of a vesicle its equilibrium shape
changes. Because the curvature dependence of the surface free energy
of the solutes cannot be written in the Helfrich form their effect
cannot be described by a simple rescaling of the bending rigidity
coefficients. We have quantitatively computed the shape changes and
the shift of the prolate-oblate transition line in the phase diagram.

In order to decide whether these effects are large enough to be
observable in experiments an estimate for the quantity $x=\rho_b
L^3/\beta\kappa$ is needed. The bending rigidity is typically of the
order of $10^{-19}\,\rm J$ \cite{Doebereiner:97}. If the rod density
$\rho_b$ is too large the rod-rod interaction becomes important which
might screen the interesting effects \cite{rods}. Since these
interactions scale with the second virial coefficient $B_2\sim DL^2$ a
useful dimensionless measure for their strength is $\rhos_b=\rho_b
DL^2$. So even if $\rhos_b$ must be limited to a small number, 0.1
say, $x$ can in principle be made arbitrarily large by choosing large
enough aspect ratios $L/D$. In practice, however, one would need
$L/D\sim O(10^3)$ to obtain $x\sim O(10)$. This is much higher than
for the ``classical'' rod-like colloidal particles like the tobacco
mosaic virus \cite{Vroege:92} but may be achievable with microtubules
for which $D=25\,\rm nm$ and $L$ can be tens of micrometers. Rather
large vesicles would have to be used so that the curvature radii are
still large compared to $L$ which justifies the neglection of higher
order terms in the curvature.

Microtubules and many other mesoscopic rod-like particles are usually
polydisperse. Thus a useful extension of the present work would be the
inclusion of polydispersity which poses no fundamental technical
problems as long as the interparticle interactions can still be
neglected. A generalization to soft particle-wall interactions seems
to be more difficult as no simple analytical expressions for more
realistic potentials exist.

Finally we mention that contrary to what is claimed in
\cref{Yaman:97:PRL} the corresponding problem for disks instead of
rods is not completely equivalent. It is easy to convince oneself
with a coin and a cup that there are configurations where a disk
touches the inside of a cylinder at two isolated points whose distance
is smaller than the disk diameter. The
orientational constraints due to these
configurations obviously cannot be described by replacing the disk
with an equivalent rod.

\section*{Acknowledgments}

I thank U. Seifert, A. Hanke, and S. Dietrich for helpful discussions
and B. Mulder and M. Bates for a critical reading of the manuscript.
This work is part of the research program of the Stichting voor
Fundamenteel Onderzoek der Materie (Foundation for Fundamental Research on
Matter) and was made possible by financial support from the Nederlandse
Organisatie voor Wetenschappelijk Onderzoek (Netherlands Organization for
the Advancement of Research). I acknowledge the financial support
of the EU through the Marie Curie TMR Fellowship programme.


\end{document}